\documentclass[doublecol,linenumbers]{epl2} % for 2 columns style with line numbers
\usepackage{bm}
\usepackage{graphicx}
\usepackage{mathptmx}      % use Times fonts if available on your TeX system

\usepackage{amsmath}

\usepackage{amssymb}

\usepackage{epstopdf}

\usepackage{subfigure}

\usepackage{subfig}

\title{L\'{e}vy walk with multiple internal states}
\shorttitle{Title} %Insert here a short version of the title if it exceeds 70 characters

\author{Pengbo Xu\inst{1} \and Weihua Deng\inst{1}}
\shortauthor{P.B. Xu and W.H. Deng}%\etal}

\institute{
  \inst{1} School of Mathematics and Statistics, Gansu Key Laboratory of Applied Mathematics and Complex Systems, Lanzhou University, Lanzhou 730000, P.R. China\\
}

\pacs{5.10.Gg}{Stochastic analysis methods}

\abstract{L\'{e}vy walk is a fundamental model with applications ranging from quantum physics to paths of animal foraging. Taking animal foraging as an example, a natural idea that comes to one's mind is to introduce the multiple internal states for dealing with the dependence of the PDF of waiting time on the energy of the animal and richness of the food at a particular location, etc; the framework can also be used to model the moving trajectories of smart animals without returning to the directions or locations which they come from immediately. After building the L\'{e}vy walk model with multiple internal states and deriving the governing equation of the distribution of the  positions of the particles, some applications are discussed with specific transition matrices. The type of diffusion for non-immediately-repeating L\'{e}vy walk is uncovered, and the distribution and average of first passage time are numerically simulated.
}

\begin{document}

\maketitle

\section{Introduction}
Diffusion is the net movement of particles from the region of high concentration to the region of low concentration. Because of the central limit theorem (CLT) \cite{Durrett:2004}, normal diffusion is very useful and well-known. In recent decades, it is found that CLT does not hold again in many natural phenomenon \cite{golding:2006,scalas:2006,zaslavsky:2002}, leading to the notion of so-called anomalous diffusion. The type of diffusion is generally classified according to its mean square displacement (MSD)\cite{metzler:2000,klafter:2011}. Specifically, for a stochastic process $x(t)$ with $\big<x^2(t) -x(0)\big>\sim t^{\alpha}$, it is respectively called normal diffusion, subdiffusion, and superdiffusion for $\alpha=1$, $0<\alpha<1$, and $\alpha>1$.

%Nowadays anomalous diffusion can be found in many natural phenomenon \cite{golding:2006,scalas:2006}. The anomalous diffusion is always classified according to its mean square displacement (MSD)\cite{metzler:2000,klafter:2011}. Specifically, for the stochastic process $x(t)$ there exists $\big<x^2(t)\big>\sim t^{\alpha}$. If $0<\alpha<1$ the process moves slower so called subdiffusion while $\alpha>1$ is superdiffusion which means the process moves faster than the normal one ($\alpha=1$).

One of the most powerful models to describe diffusion is continuous time random walk (CTRW) \cite{klafter:2011}, which always consists of two random variables saying waiting time $\tau$ and jump length $\xi$, i.e., $\tau$ is for the waiting time of each step and $\xi$ the jump length. If both the means of $\tau$ and $\xi^2$ are finite, the CTRW model describes normal diffusion, while the mean(s) of $\tau$ and/or $\xi^2$ diverge(s), it almost always characterizes anomalous diffusion except a very particular case. In fact, for the anomalous diffusion, if the mean of $\tau$ is unbounded, generally its distribution is power-law distribution $1/\tau^{1+\alpha}$ with $0<\alpha<1$; and for the divergent average of $\xi^2$, its distribution usually is $1/\xi^{1+\beta}$ with $0<\beta<2$. If $\tau$ is with exponential distribution and $\xi$ the power-law distribution, the CTRW model describes L\'evy flight \cite{metzler:2000,klafter:2011}, having divergent MSD; while L\'evy flight has wide applications \cite{mandelbrot:1982}, it also has a particular drawback of failing to be characterized by the second moment \cite{zaburdaev:2015}.

L\'evy walk (LW) is a model, which remedies the problem of divergent moments. It couples the displacement of the step and the time taking to the corresponding distance (put larger time cost to longer distance \cite{klafter:2011}); that is, the jump length and waiting time have the joint distribution $\frac{1}{2} \delta(|x|-v_0t) \phi(t)$, where $\delta$ is the Dirac function,  $\phi(t)$ is the probability density function (PDF) of the time cost of a step, and $v_0$ is a constant velocity \cite{Shlesinger:1986,shlesinger:1982,klafter:1994}. All moments of the distribution of the particle's position of LW are finite. Three different models are introduced for the two dimensional LW \cite{zaburdaev:2016}; the first one is obtained out of the one-dimensional LW by assuming that the motions along each axis, $x$ and $y$, are identical and independent one-dimensional LW processes; the second one is to allow a particle to move only along one of the axes at a time; and the third one is to allow a particle to move along all the directions uniformly; the MSDs of the three models are the same, so a generalized Pearson coefficient (PC) ${\rm PC}(t)= \langle x^2(t) y^2(t) \rangle/\langle x^2(t) \rangle\langle y^2(t) \rangle$ is introduced to distinguish the models.

Sometimes, a stochastic (physical) process has multiple internal states with applications ranging from electronic burst noise to ionic currents in cell membranes \cite{Godec:2017,Pollak:1993}. A natural idea is to introduce multiple internal states to LW models; as for its applications, for example, in animal foraging the PDF of waiting time may relate to the energy of the animal and richness of the food at a particular location \cite{xu:2017}. In this letter, we first build the LW model with multiple internal states and derive the corresponding governing equation of the PDF of positions of the particles, then focus on its applications with specific transition matrices, representing different kinds of none-immediately-repeating LWs. We find that: 1. if the LW displays superdiffusion, the none-immediately-repeating request has no influence on the MSD and the generalized PC, which is a kind of stable property of LW; 2. however, if the LW shows normal diffusion, the influence of none-immediately-repeating request emerges; 3. first passage time can be used to distinguish the processes with different transition matrices while MSD and generalized PC are always the same. We numerically simulate the distribution and average of the first passage time of LW and the relationship between its average and the considered domain.

\section{L\'{e}vy walk with multiple internal states}
Similar to \cite{xu:2017}, in this letter we also use `bra-ket' notations. That is the bras $\big<\cdot\big|$ and kets $\big|\cdot\big>$ denote the row and column vectors, respectively. L\'{e}vy walk with multiple internal states contains several different distributions of waiting time and velocity. In this letter we consider the case of finite internal states and denote the number of internal states as $N$. The distributions of waiting time and velocity are denoted as $\phi^{(i)}(\tau)$ and $h^{(i)}(\mathbf{v})$, $i=1,\ldots,N$. The particle starts its movement by randomly choosing one pair of the waiting time and velocity distributions according to the initial distribution denoted as $\big<{\rm init}\big|=(\xi_1,\ldots,\xi_N)$. Then the particle chooses its internal states with respect to the transition matrix $M$ which consists of $m_{ij}$ representing the probability of transition from the $i$-th internal state to the $j$-th one. Specifically, after one movement the particle stays, say, at the $i$-th internal state and the $i$-th row of $M$ will be the new distribution of the internal state which in the next step the particle will respect to. Assuming that the particle starts at the origin, then we have the equation of $q^{(i)}(\mathbf{r},t)$, the PDF of the particle just arriving at position $\mathbf{r}$ and $i$-th internal state at time t,
\begin{equation*}
\begin{split}
 q^{(i)}(\mathbf{r},t)=&\sum_{j=1}^{N}\int_{0}^{t}d\tau\int d\mathbf{v}m_{ji}\phi^{(j)}(\tau)h^{(j)}(\mathbf{v})q^{(j)}(\mathbf{r}-\mathbf{v}\tau,t-\tau)  \\
     & +\xi_i\delta(\mathbf{r})\delta(t).
\end{split}
\end{equation*}
Denoting $\big|q(\mathbf{r},t)\big>=(q^{(1)}(\mathbf{r},t),\ldots,q^{(N)}(\mathbf{r},t))^T$, $\Phi(\tau)={\rm diag}(\phi^{(1)}(\tau),\ldots,\phi^{(N)}(\tau))$, and $H(\mathbf{v})={\rm diag}(h^{(1)}(\mathbf{v}),\ldots,h^{(N)}(\mathbf{v}))$ leads to
\begin{equation}\label{sec2eq1}
\begin{split}
   \big|q(\mathbf{r},t)\big>= & \int_{0}^{t}d\tau\int d\mathbf{v}M^T\Phi(\tau)H(\mathbf{v})\big|q(\mathbf{r}-\mathbf{v}\tau,t-\tau)\big> \\
     &+\delta(\mathbf{r})\delta(t)\big|{\rm init}\big>.
\end{split}
\end{equation}
Similarly, we can get the equation for $\big|P(\mathbf{r},t)\big>$ consisting of $P^{(i)}(\mathbf{r},t)$, which represents the probability of the particle arriving at position $\mathbf{r}$ with the $i$-th internal state at time $t$,
\begin{equation}\label{sec2eq2}
  \big|P(\mathbf{r},t)\big>=\int_{0}^{t}d\tau\int d\mathbf{v}\Psi(\tau)H(\mathbf{v})\big|q(\mathbf{r}-\mathbf{v}\tau,\tau)\big>
\end{equation}
with $\Psi(\tau)=I-\int_{0}^{\tau}\Phi(t')dt'$. After performing Laplace and Fourier transforms, and denoting $f(s)=\int_{0}^{\infty} e^{-st}f(t)dt$ and $g(\mathbf{k})=\int e^{-i \mathbf{k}\mathbf{r}}g(\mathbf{r})d\mathbf{r}$, we have
\begin{equation}\label{sec2eq3}
\begin{split}
  \big|P(\mathbf{k},s)\big> =&\int d\mathbf{v} H(\mathbf{v})\Psi(s+i\mathbf{k}\mathbf{v}) \\
     &\cdot\bigg[I-\int M^T\Phi(s+i\mathbf{k}\mathbf{v})H(\mathbf{v}) d\mathbf{v}\bigg]^{-1}\big|{\rm init}\big>.
\end{split}
\end{equation}
If the L\'{e}vy walk just has one internal state, then eq. \eqref{sec2eq3} reduces to
\begin{equation}\label{sec2eq4}
  P(\mathbf{k},s)=\frac{\int d\mathbf{v}h(\mathbf{v})\psi(s+i\mathbf{kv})}{1-\int d\mathbf{v}\phi(s+i\mathbf{kv})h(\mathbf{v})},
\end{equation}
being the same as the equation given in \cite{zaburdaev:2015,zaburdaev:2016}. Furthermore, one can get the PDF $P(\mathbf{r},t)$ by calculating $P(\mathbf{r},t)=\big<\Sigma\big|P(\mathbf{r},t)\big>$, where $\big<\Sigma\big|$ represents the row vector that all the elements are 1.

For convenience, we use the notation $\{\{m_{11},m_{12},m_{13},m_{14}\},\\\{m_{21},
m_{22},m_{23},m_{24}\},\{m_{31},m_{32},m_{33},m_{34}\},\{m_{41},m_{42},m_{43},\\
m_{44}\}\}$ to represent the matrix
\begin{equation*}
\begin{pmatrix}
m_{11} & m_{12} & m_{13} & m_{14}\\
m_{21} & m_{22} & m_{23} & m_{24}\\
m_{31} & m_{32} & m_{33} & m_{34}\\
m_{41} & m_{42} & m_{43} & m_{44}\\
\end{pmatrix}.
\end{equation*}
And we mainly consider 7 representative transition matrices:\\
$M_0=\{\{1/4,1/4,1/4,1/4\},\{1/4,1/4,1/4,1/4\},\{1/4,1/4,\\
1/4,1/4\},\{1/4,1/4,1/4,1/4\}\}$,\\
$M_1=\{\{1/3,1/3,1/3,0\},\{1/3,1/3,0,1/3\},\{1/3,0,1/3,\\
1/3\},\{0,1/3,1/3,1/3\}\}$,\\
$M_2=\{\{0,1/2,1/2,0\},\{1/2,0,0,1/2\},\{1/2,0,0,1/2\},\{0,\\
1/2,1/2,0\}\}$,\\
$M_3=\{\{1/2,1/2,0,0\},\{0,1/2,0,1/2\},\{1/2,0,1/2,0\},\{0,0,\\
1/2,1/2\}\}$,\\
$M_4=\{\{1,0,0,0\},\{0,1,0,0\},\{0,0,1,0\},\{0,0,0,1\}\}$,\\
$M_5=\{\{0,0,1,0\},\{1,0,0,0\},\{0,0,0,1\},\{0,1,0,0\}\}$,\\
$M_6=\{\{1/3,1/3,1/3,0\},\{1/3,1/3,0,1/3\},\{1/3,0,1/3,\\
1/3\},\{1/4,1/4,1/4,1/4\}\}$.

\section{None-immediately-repeating L\'{e}vy walks}
\begin{figure}
\onefigure[width=6cm]{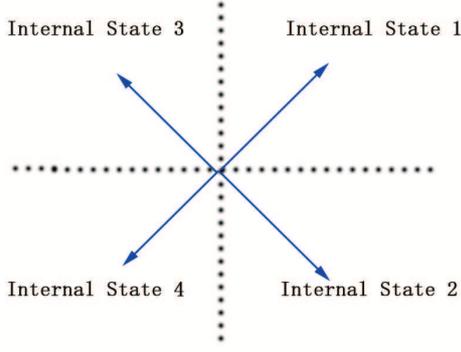}
\caption{Four different internal sates of L\'{e}vy walk. The four internal states correspond to the four different choices of the next step. Specifically, after one movement of the particle, it may choose to go to the upper right area or others according to the current position and the transition matrix. This `upper right area' is the first internal state. The other areas and internal states are defined in the same way.}
\label{fig.1}
\end{figure}

Now, we present the applications of the LW models with multiple internal states. It can be effectively used to deal with the LW with the request of not allowing to return to the direction or area that it immediately comes from. This should naturally be an intelligent animal's walk. For a particle moving in two dimensional space divided into four quadrants (see Fig. \ref{fig.1}), if taking the current position of the particle as origin, in the next step, the particle will enter one of the quadrants according to its current position and the transition matrix. The four different choices of the next step are the four internal states mentioned above.

%In this section, we mainly discuss L\'{e}vy walks which can't immediately return to the direction or area where it comes from in the previous step. This is more natural for an intelligent animal's walk. If we only consider a particle moves freely, then its next step can have 4 different choices and these choices are the internal states we mentioned above. More specifically, we show in Fig. \ref{fig.1}.
According to the internal states defined above, we can separate the distribution of velocity into four different cases and gather them to form a diagonal matrix. In this letter, we consider the uniform one of the three different types of models proposed in \cite{zaburdaev:2016}, namely, the model of particle moving along all directions uniformly. The magnitude of velocity is taken as a given constant $v_0$, and the directional angle is uniformly distributed in the interval $[0,2 \pi]$.
%In \cite{zaburdaev:2016}, 3 different kinds of L\'{e}vy walk models %in 2 dimension are given. In this letter we mainly consider the %uniform model, that is the velocity of the particle is a given %constant $v_0$ and the angel of next movement is uniformly %distributed in the interval $[0,2\pi]$.
Then one can get the diagonal matrix of the velocity $\mathbf{v}=(v_x,v_y)$ for none-immediately-repeating L\'{e}vy walk of the uniform model $H_{uniform}(\mathbf{v})={\rm diag}\big(2/(\pi v_0)\delta\big(\sqrt{v_x^2+v_y^2}-v_0\big)\kappa(v_x)\kappa(v_y), 2/(\pi v_0)\delta\big(\sqrt{v_x^2+v_y^2}-v_0\big)\kappa(v_x)\kappa(-v_y),\\
 2/(\pi v_0)\delta\big(\sqrt{v_x^2+v_y^2}-v_0\big)\kappa(-v_x)\kappa(v_y), 2/(\pi v_0)\delta\big(\sqrt{v_x^2+v_y^2}-v_0\big)\kappa(-v_x)\kappa(-v_y)\big),$ where $
\kappa(x)=\begin{cases}
1 & \text{$x>0$}\\
0 & \text{$x\leq0$}
\end{cases}.
$
Now one can see that if we choose $M_0$ and $M_1$ as transition matrices then the L\'{e}vy walk would be repeatable and have $1$ area (the area where the particle comes from in the previous step) none-immediately-repeating, respectively. The transition matrices $M_2$ and $M_3$ are for $2$ areas none-reaching L\'{e}vy walk, and $M_4$, $M_5$ are the transition matrices of the $3$ areas non-reaching walk. For the specific way of the movement of the particle with these transition matrices, one can see in \cite{xu:2017}.

To simply propose the applications of the built model, we only consider the case that all the waiting time distributions are the same. That is $\phi^{(1)}(\tau)=\cdots=\phi^{(4)}(\tau)=\gamma/(\tau_0(1+\tau/\tau_0)^{1+\gamma})$, where $\tau_0$, $\gamma>0$. For the parameters, we consider three cases: $1>\gamma>0$, $2>\gamma>1$, and $\gamma>2$. When $\gamma=1,2$, the waiting time distribution is different \cite{zumofen:1993}. Then after Laplace transform, one have
\begin{equation}\label{sec2eq5}
\begin{split}
     \Phi(s)\sim \left[1-\frac{\tau_0}{\gamma-1}s-\tau_0^\gamma \Gamma(1-\gamma)s^\gamma+\frac{\tau_0^2}{(\gamma-2)(\gamma-1)}s^2\right]{\rm I}.
\end{split}
\end{equation}
Furthermore the surviving probability has the form
\begin{equation}\label{sec2eq6}
  \begin{split}
     \Psi(s) & ={\rm diag}(\psi^{(1)}(s),\cdots,\psi^{(4)}(s))=\frac{1-\Phi(s)}{s}\\
       & =\left[\frac{\tau_0}{\gamma-1}+\tau_0^\gamma\Gamma(1-\gamma)s^{\gamma-1}-\frac{\tau_0^2}{(\gamma-2)(\gamma-1)}s\right]{\rm I}.
  \end{split}
\end{equation}
Utilizing Eq. \eqref{sec2eq5}, Eq. \eqref{sec2eq6}, the diagonal matrix $H_{uniform}(\mathbf{v})$, and noticing that $\int M^T \Phi(s+i\mathbf{k}\mathbf{v})H(\mathbf{v})d\mathbf{v}=M^T\int\Phi(s+i\mathbf{k}\mathbf{v})H(\mathbf{v})d\mathbf{v}$, we can consider each internal state separately, that is for the first internal state
\begin{equation*}
\begin{split}
&\int_{-\infty}^{\infty}dv_x \int_{-\infty}^{\infty}dv_y h^{(1)}(v_x,v_y)\phi^{(1)}(s+ik_xv_x+ik_yv_y)\\
=&\int_{-\infty}^{\infty}dv_x \int_{-\infty}^{\infty}dv_y \frac{2}{\pi v_0}\delta(\sqrt{v_x^2+v_y^2}-v_0)\kappa(v_x)\kappa(v_y)\\
&\phi^{(1)}(s+ik_xv_x+ik_yv_y)\\
=&\int_{0}^{\infty}d\rho\int_{0}^{\frac{\pi}{2}}d\theta\frac{2\rho}{\pi v_0}\delta(\rho-v_0)\phi^{(1)}(s+ik_x\rho\cos\theta+ik_y\rho\sin\theta)\\
=&\frac{2}{\pi}\int_{0}^{\frac{\pi}{2}} 1-\frac{\tau_0}{\gamma-1}(s+ik_xv_0\cos\theta+ik_yv_0\sin\theta)\\
%=&\frac{2}{\pi}\int_{0}^{\frac{\pi}{2}}1-\frac{\tau_0}{\gamma-1}(s+ik_xv_0\cos\theta+ik_yv_0\sin\theta)\\
&-\tau_0^\gamma\Gamma(1-\gamma)(s+ik_xv_0\cos\theta+ik_yv_0\sin\theta)^\gamma\\
&+\frac{\tau_0^2}{(\gamma-2)(\gamma-1)}(s+ik_xv_0\cos\theta+ik_yv_0\sin\theta)^2 d\theta.
\end{split}
\end{equation*}
Similarly, one can obtain the ones for the other internal states by simply changing the integral interval to $[-\frac{\pi}{2},0]$ for internal state $2$, $[\frac{\pi}{2},\pi]$ for internal state $3$, and $[-\pi,-\frac{\pi}{2}]$ for internal state $4$. Besides one can also get $\int d\mathbf{v} H(\mathbf{v})\Psi(s+i\mathbf{k}\mathbf{v})$ for each internal state with the same method. Then, in the Fourier-Laplace space, we get the expression of $P(kx,ky,s)=\big<\Sigma|P(kx,ky,s)\big>$ by utilizing Eq. \eqref{sec2eq3}. To obtain the MSD, one can further do the Taylor expansion at  $k_x=0$ and $k_y=0$.
%Then we use Taylor series at $k_x=0$ and $k_y=0$. Finally in the Fourier-Laplace space, we have the expression of $P(kx,ky,s)=\big<\Sigma|P(kx,ky,s)\big>$ by utilizing Eq. \eqref{sec2eq3}. 
In this letter, we consider the MSD along $X$-axis, $\big<x^2(t)\big>=\mathcal{L}^{-1}\big\{ \big(-\frac{\partial^2}{\partial k_x^2}P(k_x,k_y=0,s)\big)\big|_{k_x=0}\big\}$. By utilizing transition matrix $M_0$ (the repeatable L\'{e}vy walk) we can obtain the corresponding MSD along $X$-axis satisfying: 1) $\big<x^2(t)\big>\sim\frac{(1-\gamma)v_0^2}{2}t^2$ for $0<\gamma<1$; 2)  $\big<x^2(t)\big>\sim\frac{v_0^2\tau_0^{\gamma-1}(\gamma-1)}{(3-\gamma)(2-\gamma)}t^{3-\gamma}$ for $1<\gamma<2$; 3) $\big<x^2(t)\big>\sim \frac{v_0^2\tau_0}{\gamma-2}t$ for $\gamma>2$. These results are the same as \cite{zaburdaev:2016}. %在参考文献4的SUPPLEMENTAL INFORMATION中有相应的结果
For the other transition matrices, $M_1,\cdots,M_5$, one can also obtain the corresponding MSD shown in Tab. \ref{tab.1}.

\begin{figure}
\onefigure[width=6cm]{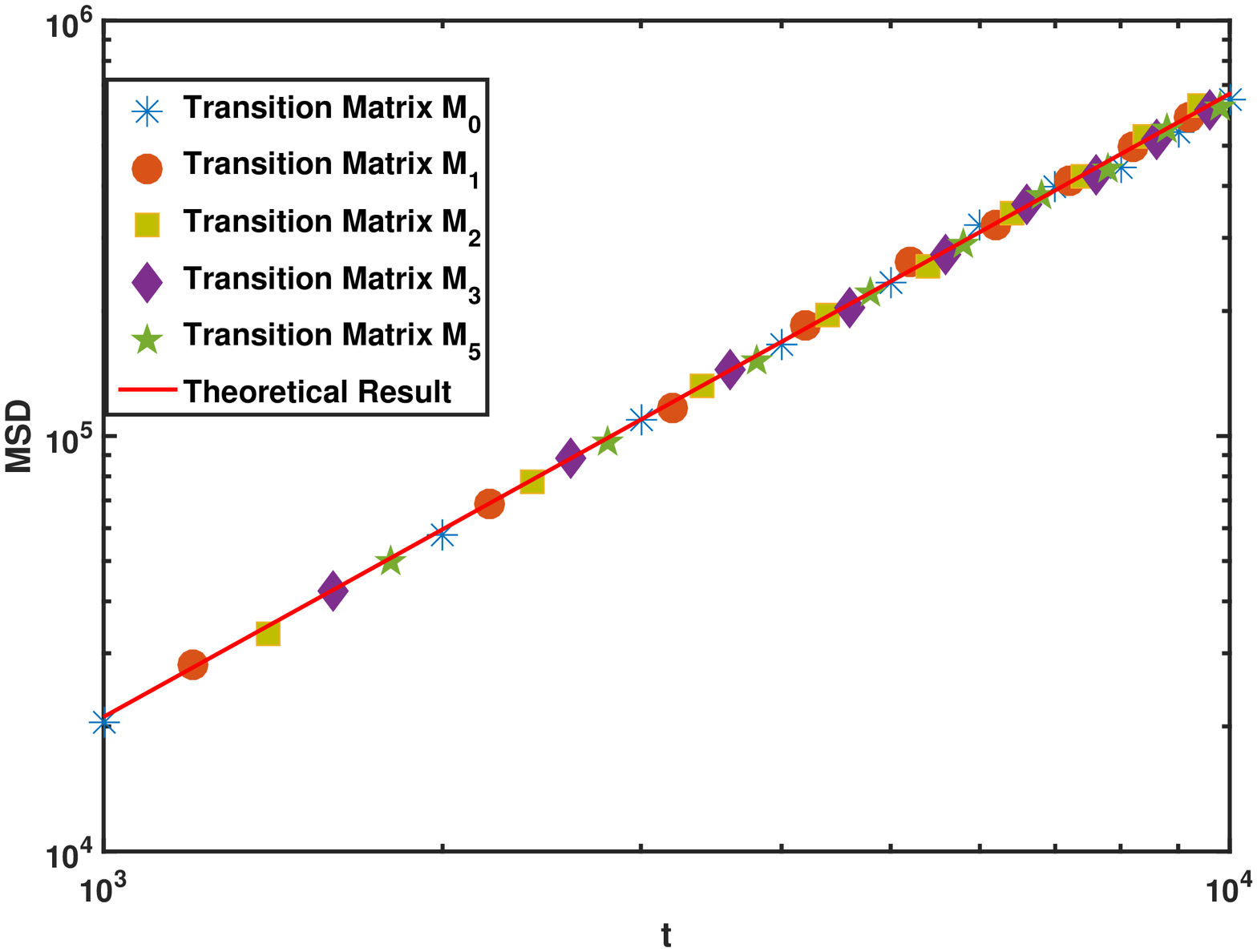}
\onefigure[width=6cm]{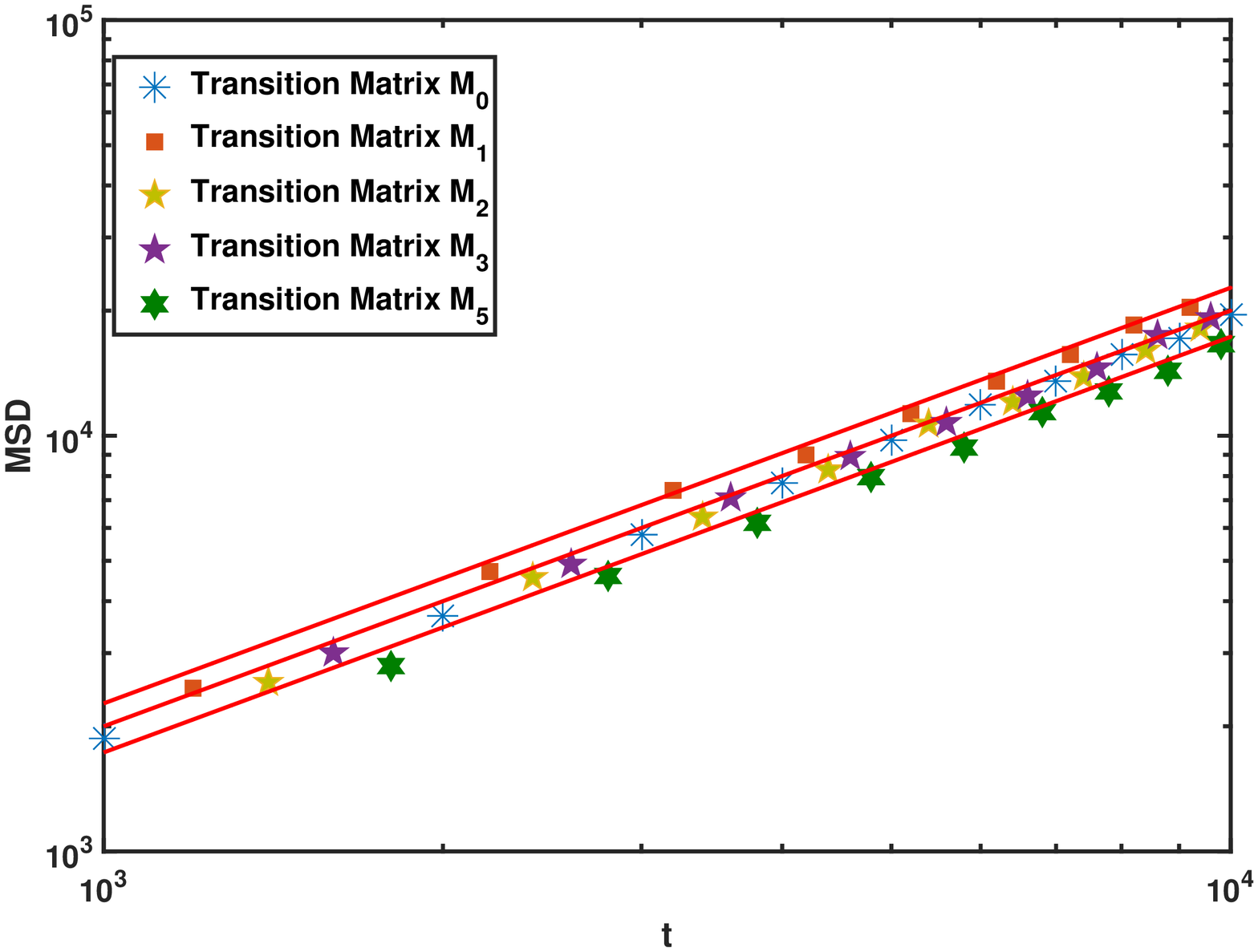}
\caption{Theoretical results and numerical ones of MSD (log-log scale), sampling over $10^4$ realizations. The upper figure is with  different transition matrices and $\gamma=1.5$, and the real line is theoretical result:  $\big<x^2(t)\big>\sim\frac{v_0^2\tau_0^{\gamma-1}(\gamma-1)}{(3-\gamma)(2-\gamma)}t^{3-\gamma}$. It further illustrates that the MSD does not depend on the transition matrix of none-immediately-repeating process (except $M_4$).
%		This figure also shows for the transition matrices of none-immediately-repeating process (except $M_4$), the MSDs always don't change.
		 In the lower figure, $\gamma=2.5$. The influence of the none-immediately-repeating begins to emerge, well verifying the theoretical results.}
\label{fig.2}
\end{figure}

\begin{table}
\caption{MSD asymptotic behaviours for different transition matrix and/or $\gamma$.}
\label{tab.1}
\begin{center}
\begin{tabular}{lccc}
     & $0<\gamma<1$                   & $1<\gamma<2$                                                                & $\gamma>2$\\
$M0$ & $\frac{(1-\gamma)v_0^2}{2}t^2$ & $\frac{v_0^2\tau_0^{\gamma-1}(\gamma-1)}{(3-\gamma)(2-\gamma)}t^{3-\gamma}$ & $\frac{v_0^2\tau_0}{\gamma-2}t$\\
$M1$ & $\frac{(1-\gamma)v_0^2}{2}t^2$ & $\frac{v_0^2\tau_0^{\gamma-1}(\gamma-1)}{(3-\gamma)(2-\gamma)}t^{3-\gamma}$ & $\frac{\tau_0 v_0^2[-8-\pi^2+\gamma(4+\pi^2)]}{(\gamma-2)(\gamma-1)\pi^2}t$\\
$M2$ & $\frac{(1-\gamma)v_0^2}{2}t^2$ & $\frac{v_0^2\tau_0^{\gamma-1}(\gamma-1)}{(3-\gamma)(2-\gamma)}t^{3-\gamma}$ & $\frac{v_0^2\tau_0}{\gamma-2}t$\\
$M3$ & $\frac{(1-\gamma)v_0^2}{2}t^2$ & $\frac{v_0^2\tau_0^{\gamma-1}(\gamma-1)}{(3-\gamma)(2-\gamma)}t^{3-\gamma}$ & $\frac{v_0^2\tau_0}{\gamma-2}t$\\
$M4$ & $\frac{v_0^2}{2}t^2$           & $\frac{\pi^2+\gamma(8-\pi^2)v_0^2}{2\pi^2}t^2$& $\frac{4}{\pi^2}v_0^2t^2$\\
$M5$ & $\frac{(1-\gamma)v_0^2}{2}t^2$ & $\frac{v_0^2\tau_0^{\gamma-1}(\gamma-1)}{(3-\gamma)(2-\gamma)}t^{3-\gamma}$ & $\frac{\tau_0 v_0^2[8-\pi^2+\gamma(-4+\pi^2)]}{(\gamma-2)(\gamma-1)\pi^2}t$\\
\end{tabular}
\end{center}
\end{table}
From Tab. \ref{tab.1}, one can clearly conclude that none-immediately-repeating has no influence on the MSD of L\'{e}vy walk for $0<\gamma<1$ and $1<\gamma<2$ (except $M_4$ representing that the L\'{e}vy walk can't change its internal state once determined by the initial distribution). The numerical simulations are presented in Fig. \ref{fig.2}, well confirming the theoretical results.
Besides, for $M_0,\cdots,M_5$ (except $M_4$) the generalized PCs \cite{zaburdaev:2016}, defined as $PC=\frac{\big<x^2(t)y^2(t)\big>}{\big<x^2(t)\big>\big<y^2(t)\big>}$,
are also calculated, being usually used to distinguish the processes if they have the same MSD. 
% which are always used to distinguish each different kind of process if they have the same MSD are also calculated \cite{zaburdaev:2016}. 
However, after calculation we find that the generalized PCs are also the same for $0<\gamma<1$ and $1<\gamma<2$, that is
\begin{equation*}
PC=\begin{cases}
\frac{2(6-3\gamma-\gamma^2)}{4!(1-\gamma)} & \text{$0<\gamma<1$}\\
\frac{(3-\gamma)^2(2-\gamma)^2}{2(5-\gamma)(4-\gamma)(\gamma-1)}(\frac{t}{\tau_0})^{\gamma-1} & \text{$1<\gamma<2$}.
\end{cases}.
\end{equation*}
In other words, the generalized PC can't distinguish these processes anymore. This is a kind of stable property of L\'{e}vy walk. But, when $\gamma>2$, it is completely different. From Tab. \ref{tab.1}, it can be seen that the transition matrix $M_1$ accelerates the  diffusion of L\'{e}vy walk; for $M_2$ and $M_3$ the diffusions of L\'{e}vy walks are neither accelerated nor decelerated; $M_4$ makes L\'{e}vy walk diffuse ballistically, and $M_5$ makes the diffusion slower.
%from Tab. \ref{tab.1} we can conclude for transition matrix $M_1$ the L\'{e}vy walk is accelerated, for $M_2$ and $M_3$ the L\'{e}vy walks are neither accelerated nor decelerated, for $M_4$ L\'{e}vy walk becomes ballistic diffusion and $M_5$ makes the process diffuse slower. 
All these effects of none-immediately-repeating for $\gamma>2$ are the same as the ones given in \cite{xu:2017}.

\begin{figure}
\centering
\subfigure{\includegraphics[width=3cm]{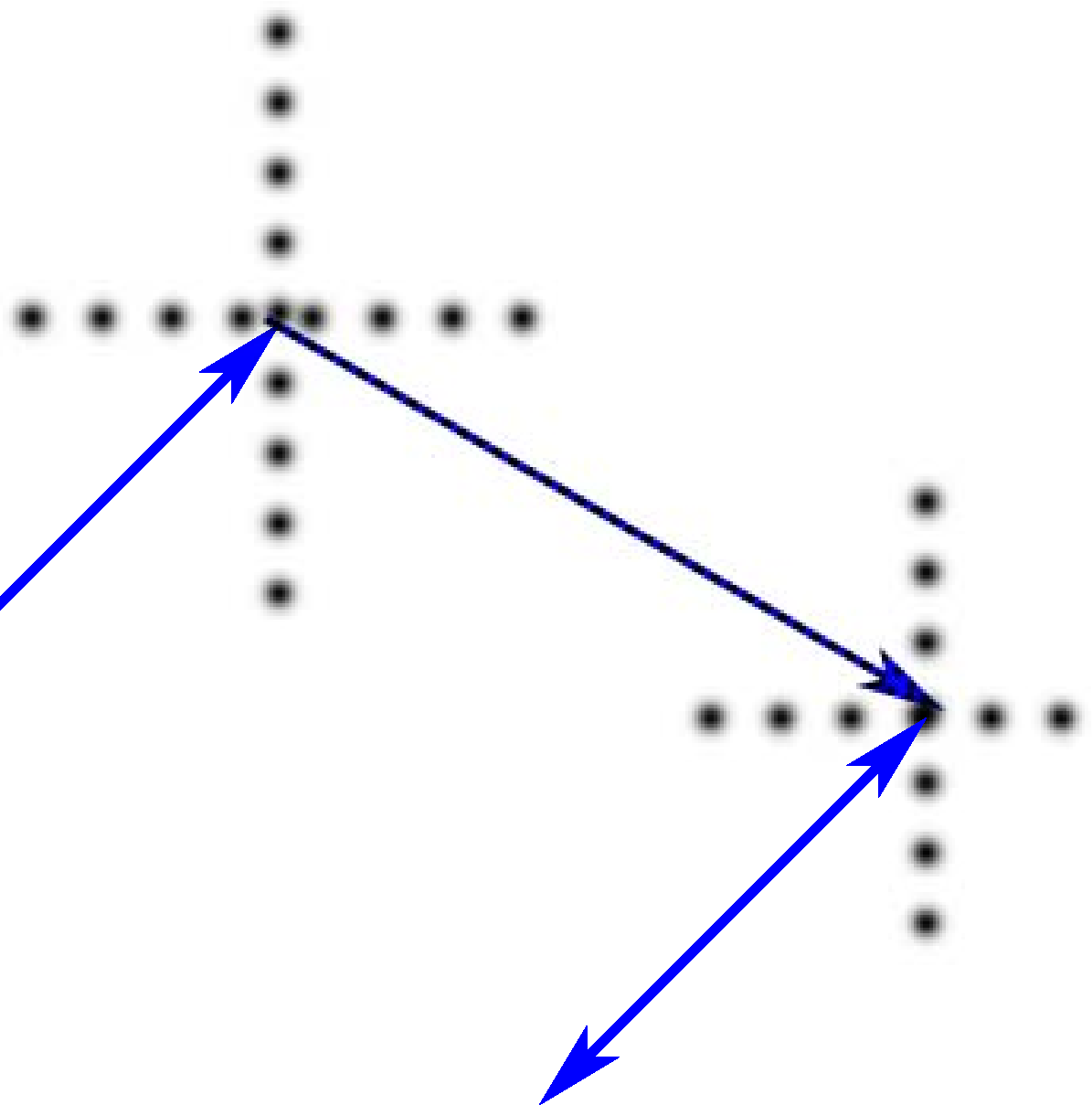}}
\subfigure{\includegraphics[width=5cm]{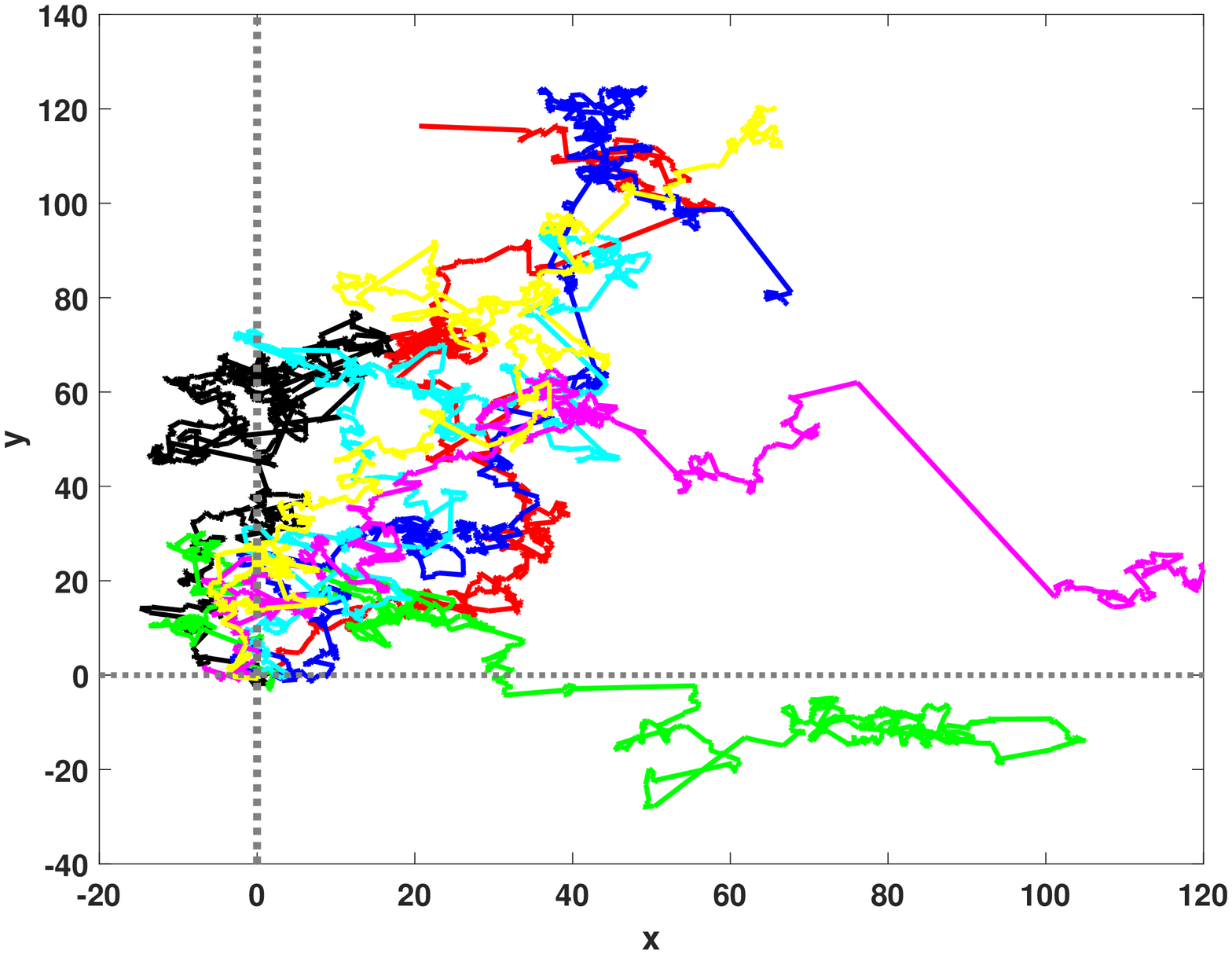}}
\caption{Trajectories of the process with transition matrix $M_6$. The left figure is for some steps of the process, in which the line with two arrows represents that the particle moves towards lower left and then goes back to upper right direction in the next step.
Some trajectories of the process with transition matrix $M_6$ are shown in the right figure ($\gamma=2.5$); it can be observed that the particles have the trend of moving towards upper right direction.	
	}
\label{fig.3}
\end{figure}

\begin{figure}
\onefigure[width=6cm]{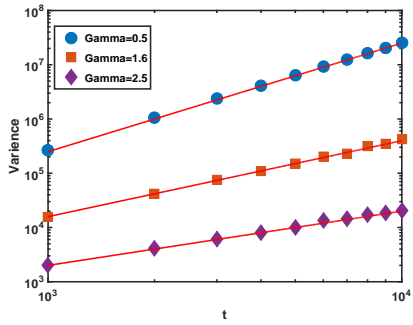}
\caption{MSD of the process with the transition matrix $M_6$ and different $\gamma$, sampling over $10^4$ realizations. The real lines in the figure are theoretical results (the upper, middle, lower real lines are with the slope of $2$, $1.4$, and $1$ respectively), being well verified by numerical results.}
\label{fig.4}
\end{figure}

Next we consider a more interesting L\'{e}vy walk that it won't return to the area of just coming from if the current internal states are $1$, $2$, $3$, but it can move freely if the particle is at the 4-th internal state.
%Next we consider the of L\'{e}vy walk won't move back to the previous area which the particle comes from when the current internal state is 1, 2, 3. However when the particle at the 4-th internal state, the particle can freely move. 
From the trajectories shown in Fig. \ref{fig.3}, the process with  transition matrix $M_6$ always has the trend of moving to the upper right direction. This is because in the transition matrix $M_6$, the probability of moving to this direction is a little bit big.
%插入模型运动图
%And in this specific case, the transition matrix is $M_6$. 
With this transition matrix, some very interesting phenomena are detected. By the same method as above, one can obtain the PDF in the Fourier-Laplace space $P(\mathbf{k},s)$. From the movement of the particle one can see that this process is not symmetric any more. Therefore we would consider the variance of the process, that is,  ${\rm Var}\big(x(t)\big)=\big<x^2(t)\big>-\big<x(t)\big>^2$. First we calculate the average of displacement by
\begin{equation*}
  \big<x(t)\big>=\mathcal{L}^{-1}\left(i\frac{\partial}{\partial k_x}P(k_x,k_y=0,s)\right)\bigg|_{k_x=0}.
\end{equation*}
After some calculations, $\big<x(t)\big>\sim\frac{4v_0}{19\pi}t$ for  sufficiently long $t$ and $0<\gamma<1$, $1<\gamma<2$, $\gamma>2$. For $\gamma>2$, we have
\begin{equation}
  \big<x^2(t)\big>\sim\frac{32 v_0^2}{2\cdot361\pi^2}t^2+C_1 t,
\end{equation}
where $C_1$ is a constant. Thus we obtain the variance ${\rm Var}\big(x(t)\big)\sim C_1t$. When $1<\gamma<2$ and $0<\gamma<1$, the variance asymptotically behaves as $C_2 t^{3-\gamma}$ and $C_3 t^2$, respectively, where $C_2$ and $C_3$ are still constants. By comparing with the variance (or MSD for symmetric process) given in \cite{zaburdaev:2015}, one can find that for the process with the transition matrix $M_6$, the exponent of variance does not change, being confirmed by numerical simulations shown in Fig. \ref{fig.4}.
If considering subdiffusion with multiple internal states \cite{xu:2017}, a completely different story happens. 
% If we consider subdiffusion with multiple internal states\cite{xu:2017}, by 
Choosing the matrix of waiting time distribution as  $\Phi(s)\sim(1-s^{\gamma})I$, where $0<\gamma<1$, and the jumping length distribution matrix shown in Eq. \eqref{AppEq2}, from the calculations in Appendix, we have
\begin{equation*}
  {\rm Var}\big(x(t)\big)\sim\frac{8}{361\pi}\left[\frac{2}{\Gamma(1+2\gamma)}-\frac{1}{\Gamma(1+\gamma)^2}\right]t^{2\gamma},
\end{equation*}
which shows that the transition matrix has a fundamental influence on the variance by totally changing the exponent from $\gamma$ to $2\gamma$. Obviously, L\'{e}vy walk is much more stable than the CTRW model in the aspect of the exponent of variance. This is also a major difference between CTRW model and L\'{e}vy walk.
%%数值模拟M_6的图
Besides, for the other models shown in \cite{zaburdaev:2016}, one can also construct the corresponding velocity and waiting time distribution matrices. The stable properties given in this letter also exist.
\section{Distribution and average of first passage time}
In this section, we mainly analyze the distribution of the first passage time by using numerical simulations. In the previous sections, we mainly discuss the stable properties of L\'{e}vy walk, which means the MSDs and the generalized PCs are always the same for the L\'{e}vy walk with multiple internal states whose transition matrices are $M_1$, $M_2$, $M_3$, $M_5$ for $0<\gamma<1$ and $1<\gamma<2$. In this section, the L\'{e}vy walk with transition matrix $M_0$ representing the repeatable case is always considered as a standard one. Now one question naturally coming into our mind  is how to distinguish these processes. Fortunately, we find that the first passage time can distinguish them very well. First passage time has many applications and ways of analysis in mathematics, physics, chemistry, and engineering \cite{benjacob:1982,bobrovsky:1982,carmeli:1983,day:1990,gao:2014,duan:2015,deng:2017,dybiec:2017}.

First we specify a domain (in our simulation, we choose a circle with  radius $r=100$ and its center is origin). Once the particle touches boundary we stop the movement and write down the corresponding time as the first passage time. One of the major differences between CTRW model and L\'{e}vy walk is that each `step' of the latter one will cost time, however the previous one does not. And this difference also affects the first passage time.
\begin{figure}
\centering
\subfigure{\includegraphics[width=6cm]{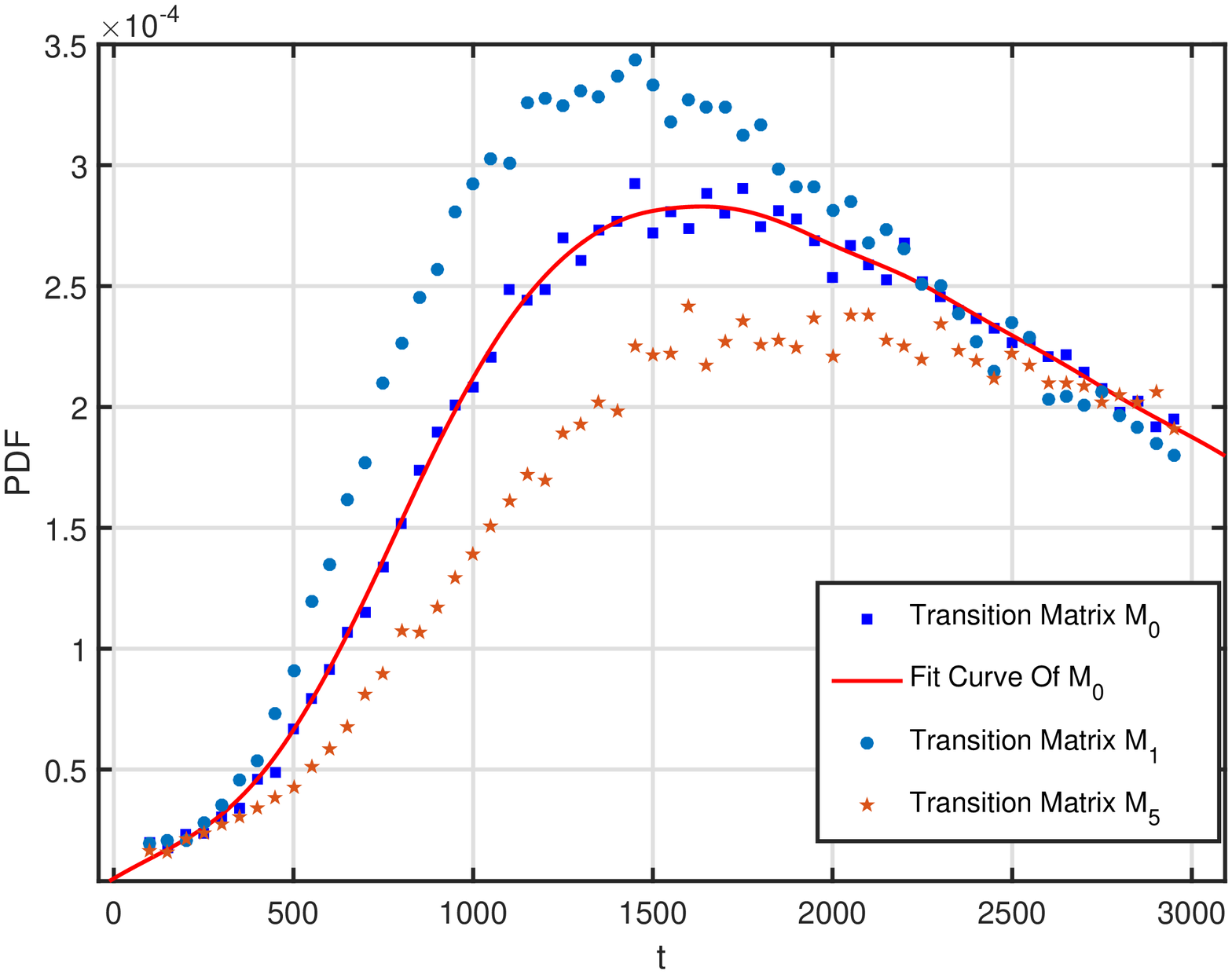}}
\subfigure{\includegraphics[width=6cm]{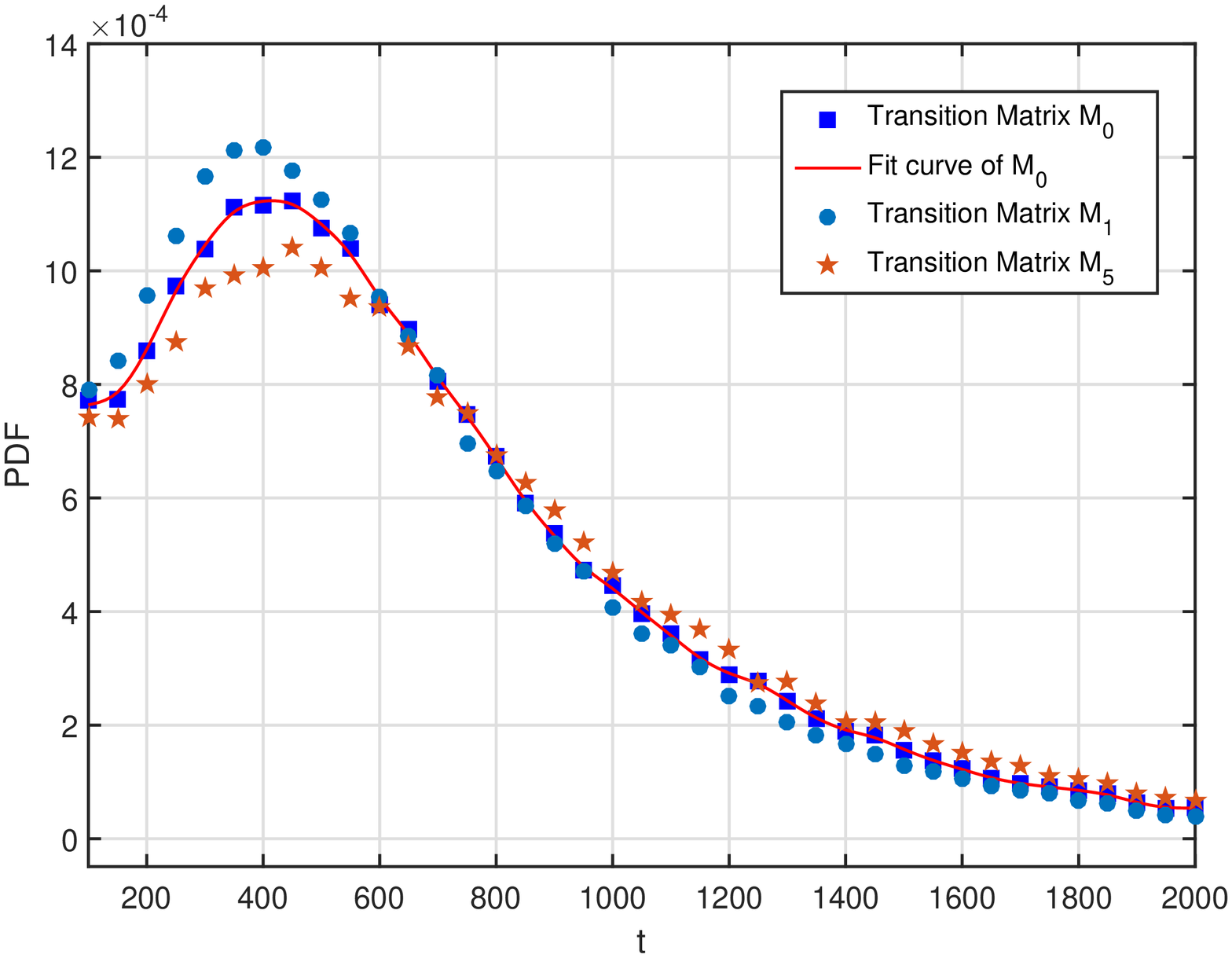}}
\caption{Distributions of the first passage time. For the above figure $\gamma=2.5$ while the lower one $\gamma=1.5$; the differences among the L\'{e}vy walks with the transition matrices $M_0$, $M_1$ and $M_5$ can be easily observed. The results of $M_2$ and $M_3$ are the same as the ones of $M_0$ (not showing in the figure). The distributions for $M_1$ in both figures are taller and thinner than the ones of $M_0$ while the ones for $M_5$ are shorter and fatter.}
\label{fig.5}
\end{figure}
For the simulation results, as shown in Fig. \ref{fig.5}, for $\gamma>2$ we can easily distinguish the processes with transition matrices $M_0$ (or $M_2$, $M_3$), $M_1$, and $M_5$. Specifically, comparing with the repeatable L\'{e}vy walk (with transition matrix $M_0$), the first passage time distribution of the process with transition matrix $M_1$ is taller and thinner, while the process with the transition matrix $M_5$ is shorter and fatter. This is might due to that the transition matrix $M_1$ makes the particle move faster while $M_5$ makes the process slower. The first passage time distribution of the L\'{e}vy walk with transition matrices $M_2$ and $M_3$ are the same as $M_0$, which means that the diffusion process is neither accelerated nor decelerated. These results are consistent with the ones shown in MSD. Next if we choose $1<\gamma<2$, from Fig. \ref{fig.5}, one can also find the same phenomenon as MSD although it is less obvious. This indicates in some sense that the diffusion process with transition matrix $M_1$ is faster while $M_5$ slower although the MSDs for $1<\gamma<2$ are the same. Finally for  $0<\gamma<1$, the first passage time distributions of the diffusion  processes with the transition matrices $M_0$, $M_1$ and $M_5$ are almost the same. However one can still notice that $M_1$ makes the distribution thinner while $M_5$ fatter.

\begin{figure}
\onefigure[width=6cm]{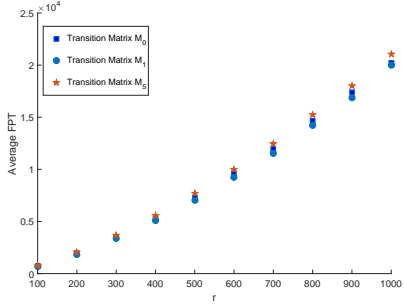}
\caption{Average of the first passage time for $\gamma=1.5$ by sampling over $10^4$ realizations. The average first passage time changing with the boundary of the domain is influenced by the transition matrix.}
\label{fig.6}
\end{figure}

From the numerical simulations of the distribution of the first passage time, one can also see that the influences of the transition matrices may appear in the average of the first passage time, as shown in Fig. \ref{fig.6}. The numerical results shown in Fig. \ref{fig.6} also turn out that the transition matrices $M_1$ and $M_5$ can accelerate and decelerate the diffusion process for $1<\gamma<2$, respectively, which are in accordance with the results of the distribution of the first passage time. For $0<\gamma<1$, we can hardly find the differences among the processes with the transition matrices $M_0$, $M_1$, and $M_5$; and the average of the first passage time, denoted as $\big<\tau\big>$, asymptotically behaves as
\begin{equation*}
  \big<\tau\big>\sim\begin{cases}
r & \text{$0<\gamma<1$},\\
r^{\gamma}& \text{$1<\gamma<2$},\\
r^2 & \text{$\gamma>2$},
\end{cases}
\end{equation*}
where $r$ is radius of the circle domain.

\section{Conclusion}  This letter builds the model of L\'{e}vy walk  with multiple internal states. The Fokker-Planck equation is obtained in Fourier-Laplace space. As applications of the model, we consider the none-immediately-repeating L\'{e}vy walks with four internal states, corresponding to $4$ by $4$ transition matrix of the model. Seven representative matrices are considered, i.e., $M_0$, $M_1$, $M_2$, $M_3$, $M_4$, $M_4$, and $M_6$.
%In this letter we mainly talk about the L\'{e}vy walk with multiple internal states. We first build up the model and obtain the equations in the Fourier-Laplace space. Basing on these equations we choose a specific problem, none-immediately-repeating L\'{e}vy walk, and give this problem four internal states. According to the internal sates we choose, we mainly study 6 different transition matrices. 
For the transition matrices $M_0$, $M_1$, $M_2$, $M_3$ and $M_5$, one can see that the MSDs and generalized PCs are always the same for $1<\gamma<2$ and $\gamma>2$. Thus we find the stable properties of L\'{e}vy walks, that is, for $1<\gamma<2$ and $\gamma>2$ if we change the L\'{e}vy walk slightly the final MSD does not change. Based on the analysis of models with transition matrix $M_6$, some more interesting phenomena are detected; in particular, we analyze the subdiffusion model, which shows that for the transition matrix $M_6$, the variance of subdiffusion asymptotically behaves as $t^{2\gamma}$ while the variance of the subdiffusion process with $M_0$ is $t^{\gamma}$. We also use first passage time and its average to further distinguish the L\'{e}vy walks with different transition matrices.

\acknowledgments
This work was supported by the National Natural Science
Foundation of China under Grant No. 11671182, and the Fundamental Research Funds for the Central Universities under Grant No. lzujbky-2017-ot10.

\section{Appendix}
\emph{Calculations of $\big<x(t)\big>$, $\big<x^2(t)\big>$ and $Var\big(x(t)\big)$ of subdiffusion with multiple internal states and transition matrix $M_6$.} From \cite{xu:2017}, we have
\begin{equation}\label{AppEq1}
\big|P(\mathbf{k},s)\big>=\frac{{I}-{\Phi}(s)}{s}\big[I-M^T H(\mathbf{k},s)\big]\big|{\rm init}\big>,
\end{equation}
where $H(\mathbf{x},t)=\Lambda(\mathbf{x})\Phi(t)$ and $\Lambda(\mathbf{x})$, $\Phi(t)$ are matrices of jump length and waiting time distribution, respectively. In this section we still consider the same internal states used in the third section. We take the jump length distribution matrix as $\Lambda(x,y)={\rm diag}\big(\gamma^{+}(x)\gamma^{+}(y), \gamma^{+}(x)\gamma^{-}(y), \gamma^{-}(x)\gamma^{+}(y), \gamma^{-}(x)\gamma^{-}(y)\big)$, where
$
\gamma^{+}(l)=\begin{cases}
\sqrt{\frac{2}{\pi\sigma^2}}\exp\left(-\frac{l^2}{2\sigma^2}\right) & \text{$l\geqslant0$}\\
0 & \text{$l<0$}
\end{cases},
$ $
\gamma^{-}(l)=\begin{cases}
0 & \text{$l\geqslant0$}\\
\sqrt{\frac{2}{\pi\sigma^2}}\exp\left(-\frac{l^2}{2\sigma^2}\right) & \text{$l<0$}
\end{cases}.
$ And the Fourier transform of $\Lambda(x,y)$ w.r.t. $x$ and $y$ is
\begin{equation}\label{AppEq2}
\begin{split}
&{\Lambda}(k_x,k_y)\\
&={\rm diag}\Bigg(\exp\left(-\frac{k_x^2+k_y^2}{2}\right)\left(1+\sqrt{\frac{2}{\pi}}i k_x\right)\left(1+\sqrt{\frac{2}{\pi}}i k_y\right),\\
&~~~~~~~~~~~~~~~\exp\left(-\frac{k_x^2+k_y^2}{2}\right)\left(1+\sqrt{\frac{2}{\pi}}i k_x\right)\left(1-\sqrt{\frac{2}{\pi}}i k_y\right),\\
&~~~~~~~~~~~~~~~\exp\left(-\frac{k_x^2+k_y^2}{2}\right)\left(1-\sqrt{\frac{2}{\pi}}i k_x\right)\left(1+\sqrt{\frac{2}{\pi}}i k_y\right),\\
&~~~~~~~~~~~~~~~\exp\left(-\frac{k_x^2+k_y^2}{2}\right)\left(1-\sqrt{\frac{2}{\pi}}i k_x\right)\left(1-\sqrt{\frac{2}{\pi}}i k_y\right)\Bigg).
\end{split}
\end{equation}
The matrix of distribution of waiting time asymptotically behaves as $\Phi(s)\sim(1-s^\gamma)I$ in Laplace space, where $0<\gamma<1$. By utilizing Eq. \eqref{AppEq1}, Eq. \eqref{AppEq2} and $\Phi(s)$, one can obtain the corresponding $\big<x(t)\big>$ and $\big<x^2(t)\big>$ with the same method in section 3, that is,
\begin{equation*}
  \big<x(t)\big>\sim\frac{2}{19}\sqrt{\frac{2}{\pi}}\frac{1}{\Gamma(1+\gamma)}t^\gamma,
\end{equation*}
and
\begin{equation*}
  \big<x^2(t)\big>\sim\frac{16}{361\pi}\frac{1}{\Gamma(1+2\gamma)}t^{2\gamma}.
\end{equation*}
Thus one obtains the variance
\begin{equation*}
  {\rm Var}\big(x(t)\big)\sim\frac{8}{361\pi}\left[\frac{2}{\Gamma(1+2\gamma)}-\frac{1}{\Gamma(1+\gamma)^2}\right]t^{2\gamma}.
\end{equation*}

\end{document}